\documentstyle[11pt,newpasp,twoside,epsf]{article} 
\def\edcomment#1{\iffalse\marginpar{\raggedright\sl#1\/}\else\relax\fi} 
\reversemarginpar 

\begin{document} 
\title{Globular Clusters in the Sgr Stream and other structures}

\author{Michele Bellazzini} 
\affil{INAF - Osservatorio Astronomico di Bologna, Via Ranzani 1, 40127, 
Bologna, Italy} 
\author{Rodrigo Ibata} 
\affil{Observatoire de Strasbourg, 67000 Strasbourg, France} 
\author{Francesco R. Ferraro} 
\affil{Dip. di Astronomia, Univ. di Bologna, Via Ranzani 1, 40127, 
Bologna, Italy} 

\begin{abstract} 
The possible association of Galactic globular clusters with the Sgr Stream is 
shortly reviewed at the light of the most recent observations of the Stream. 
$\ga 20$ \% of the Galactic globular clusters with $R_{GC}\ge 10$ Kpc are found
to lie within or very nearby to the Sgr Stream as traced over the whole sky by
the M giants from the 2MASS catalogue, in agreement with the results by
Bellazzini, Ferraro and Ibata (2003a). Another possible association of outer
halo globulars (NGC~1904, NGC~1851, NGC~2808, and NGC~2298) is also noted and 
briefly discussed.

\end{abstract}

\section{Introduction} 
It is now clear that the disruption of the Sgr dwarf spheroidal galaxy (Sgr
dSph) provided a non-negligible contribution to the assembly of the Galactic
Halo (see Majewski et al. 2003, hereafter M03, Newberg et al. 2002 
and references therein). 
Since the Sgr galaxy has its own globular cluster system it is natural to check
if the Sgr dSph has also contributed to the building up of the Galactic globular
cluster system. As a first attempt (Bellazzini et al. 2003a, hereafter Pap-I)
we searched for correlations between the
(X,Y,Z,$V_r$) phase-space distribution of the
Galactic globulars having 10 kpc $\le R_{GC} \le$ 40 kpc 
(Outer Halo globulars, OH) with the orbit and  the N-body numerical 
simulation of the Sgr dSph computed by Ibata et al. (2001). The comparison was
limited to the orbital path since present time to $\sim 1$ gyr in the past.
A significant correlation was indeed found: the considered model orbit is
a preferential subset of the phase space for OH clusters. The result strongly
suggest that some of the OH clusters is associated with the Sgr Stream but its
statistical nature prevents any firm conclusion about the individual clusters.
Nevertheless, we could rank the clusters that are candidate members of the 
Sgr Stream according to their phase-space distance from the orbit. 

Using the catalogue of the 2nd Incremental Data Release of the 2MASS survey we
have tested two of our top-rank candidates, e.g. Pal~12 and NGC~4147
(Bellazzini et al. 2003b, Pap-II). We searched an excess of M giants similar to
those found in the main body of the Sgr galaxy in the NIR color-magnitude
diagram (CMD) of wide fields around the target clusters and we found a
significant signal in both cases, showing that Pal~12 and NGC~4147 are
immersed into the Sgr Stream (see Pap-II for details and references).

\begin{figure}
\plotone{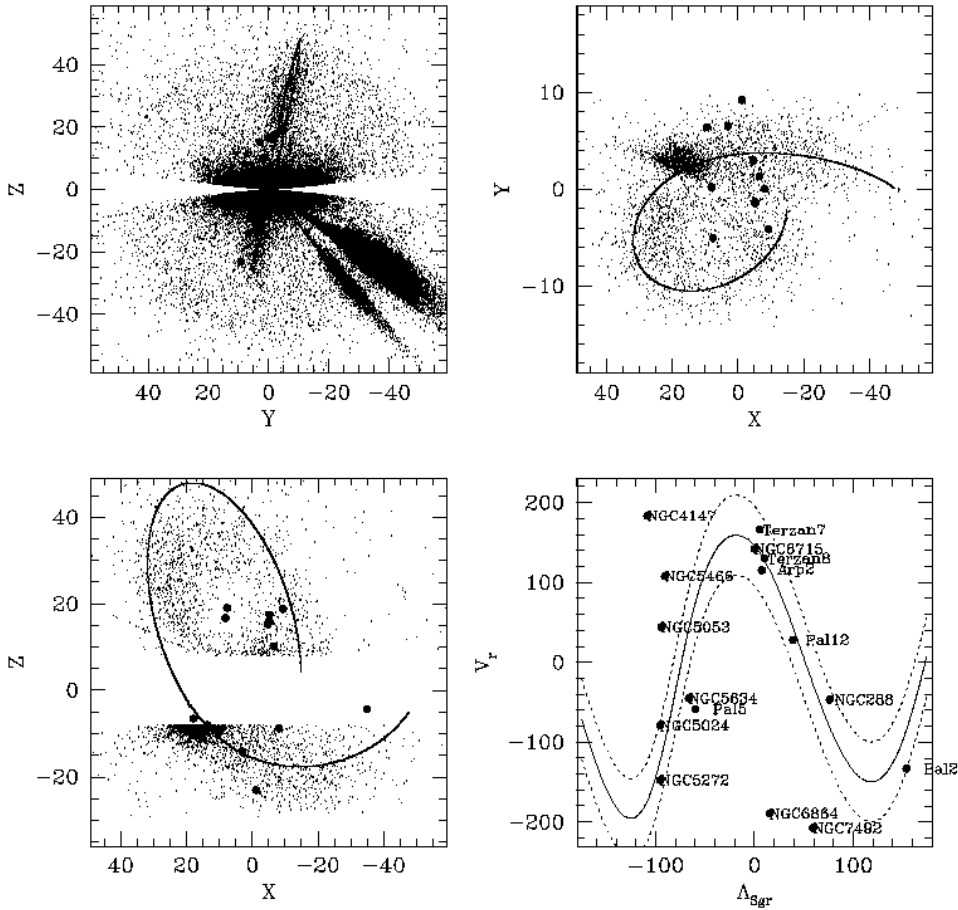}
\caption{Upper left: YZ distribution of 2MASS M giants selected as in M03. The
Sgr Stream is seen nearly edge on in this plane. The stars and clusters (filled
circles)
projected into the Stream in the YZ plane are reported in the XY and XZ planes 
(upper right and lower left). In the lower right panel the distribution of
radial velocities as a function of orbital longitude of the candidate clusters
are compared with the predictions of the model by Ibata et al.(2001).
The path of the model orbit is reported in all the plots as a continuous line.
It is quite clear that the model needs to be refined to correctly fit the
observed spatial distribution of the Sgr Stream as traced by M giants.
Excluding the known Sgr globulars (M~54, Arp~2, Ter~8 and Ter~7) there are 10
clusters that lie within or very nearby to the Sgr Stream. The measure of radial
velocities along the Stream will easily discriminate between true physical 
associations and interlopers. In any case it is clear that a few clusters must
be physically associated with the Stream. The two clusters that lie in the the
hole of the ring formed by the Sgr Stream in the XZ plane are NGC~5634 and Pal~5
which are clearly not associated with the part of the Stream that is traced by M
giants but may be related with older arms of the Stream (see Bellazzini et
al. 2002 and Vivas, this meeting).}
\end{figure}

\section{The All Sky view}

Having at disposal the All Sky data release of 2MASS, M03 have traced the Sgr
Stream over the whole sky using the M giants in a similar way to that adopted in
Pap-II. In Fig.~1 we plot the distribution of the 2MASS M giants 
(selected in the same way as M03 and using their relations for the photometric 
parallax) in different projections of the Galactocentric cartesian coordinates.
The OH clusters that matches the YZ distribution of Stream stars (the Stream is
seen nearly edge-on in this projection) have been reported as large filled
circles also in the other projections. It is quite clear that the majority of
the selected clusters fits into the Stream distribution in any projection.
Six of these clusters were also included in the list of best candidates of
Pap-I. In the lower-left panel of Fig.~1 the distribution of
radial velocities as a function of orbital longitude of the candidate clusters
are compared with the predictions of the model by Ibata et al.(2001). As evident
from the other plots the model has to be refined to correctly match the
distribution of Stream stars, however the comparison is strongly suggestive.
It is quite clear that some of the 10 clusters lying into the Stream has to be
physically associated with the Stream. The comparison with the distribution of
observed radial velocities along the Stream will soon discriminate between true
physical associations and interlopers (see Majewski and Law, these proceedings).
Therefore it seems quite firmly established that the disruption of the Sgr
galaxy had a significant role in the assembly of the Galactic globular cluster
system (see Pap-I). As far as individual clusters will pass the ``radial
velocity test'' the properties of the original globular cluster system of the
Sgr galaxy will become clearer and the early evolution of the galaxy could be
more deeply investigated.

\begin{figure}
\plottwo{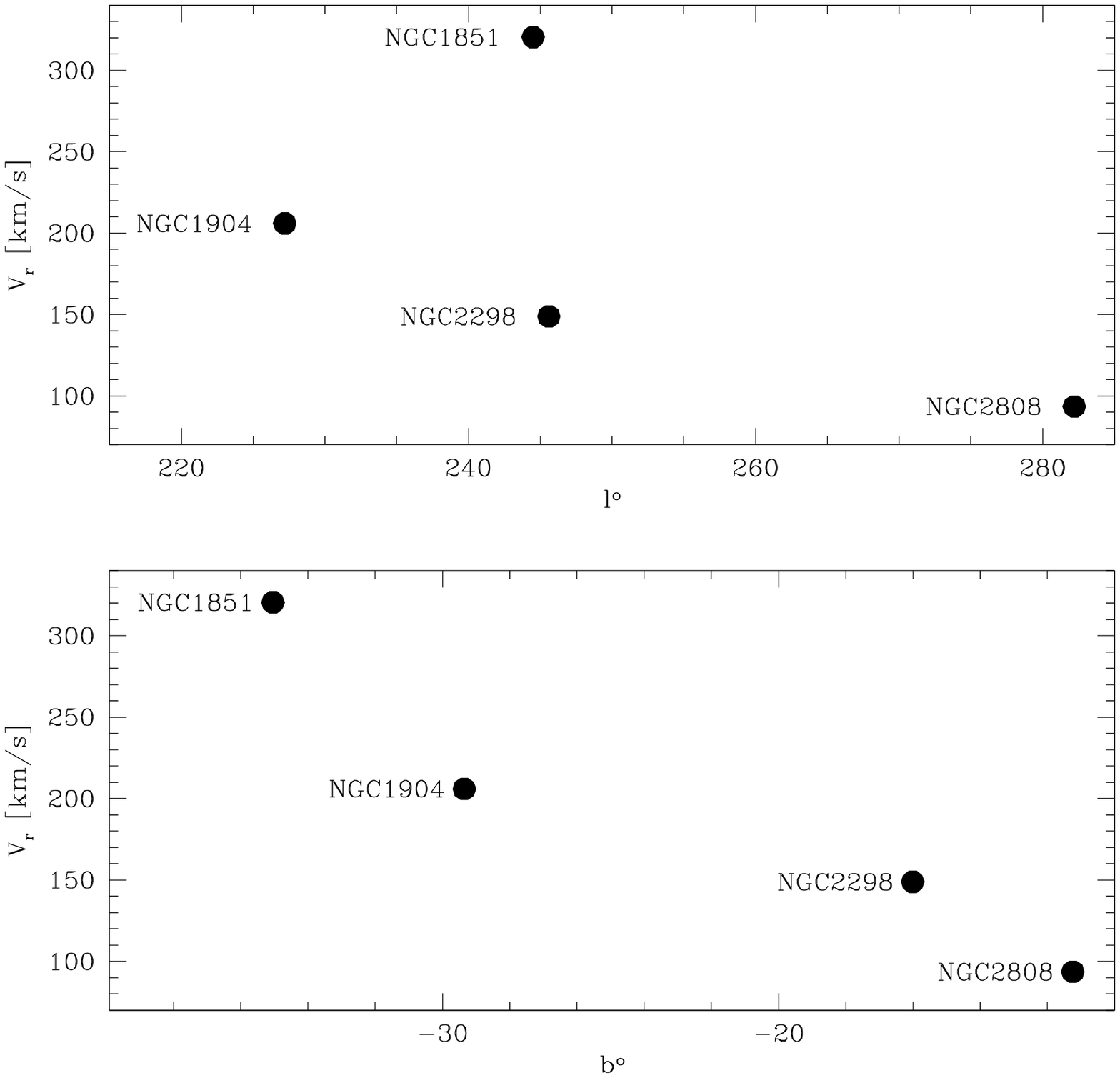}{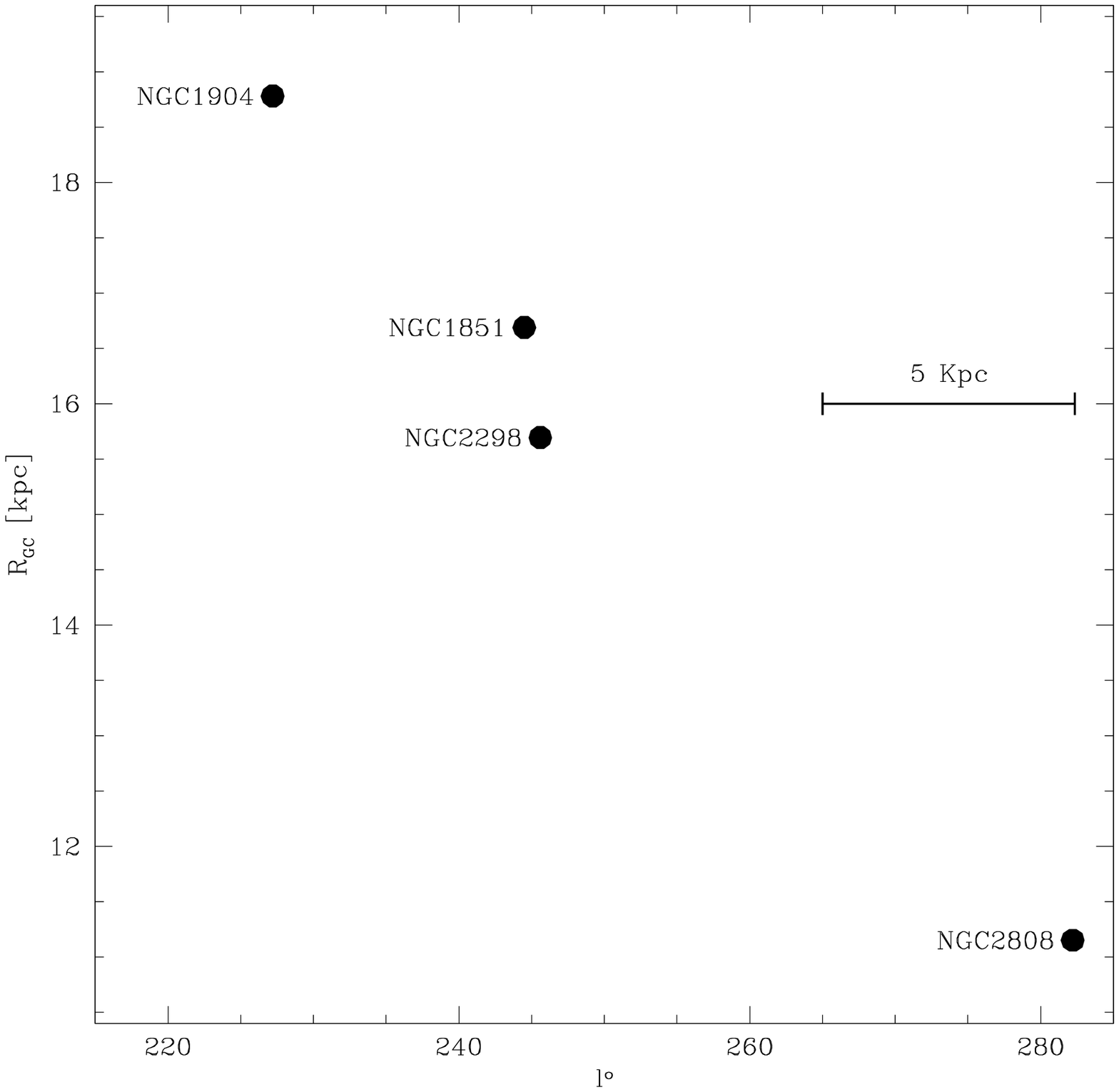}
\caption{A remarkable possible association of OH clusters not related to the Sgr
Stream. Left panel: Galactic coordinates vs. radial velocity. Right panel:
Galactic longitude vs. Galactocentric distance. The mean mutual distance
between the considered clusters is $6.2 \pm 2.7$ kpc.}
\end{figure}

\section{Other GC groups in the Outer Halo?} 

The four globular clusters associated with the main body of the Sgr galaxy
constitute a remarkable overdensity in the phase-space distribution of OH
clusters. The maximum mutual distance between any two of them is $< 6$ kpc and
the maximum difference in $V_r$ is $\simeq 50$ km/s. From a historical point of
view it is quite interesting to recall that such a structure hasn't been noted
until the Sgr galaxy was discovered. 

Is it possible that we are still
overlooking similar substructures in the globular cluster system? If candidate
Sgr Stream members are excluded (f.e. NGC~4147, NGC~5053 and NGC~5466 are quite
close one to another in the X,Y,Z,$V_r$ phase-space) there is only one 
group that shows remarkable clustering properties. The possible members of the
group are NGC~1851, NGC~1904, NGC~2298, and NGC~2808. The distances of
the latter clusters from NGC~1851 are 3.5, 4.0, and 8.0 kpc respectively.
The radial velocities of these clusters
correlates quite well with their galactic latitude and their galactocentric
distance correlates with galactic longitude (see Fig.~2, left panels). The
metallicity is also quite similar ([Fe/H]=-1.03, -1.37, -1.71, -1.11) and all 
of them have an extended blue Horizontal Branch (HB). In particular, NGC~1851
and NGC~2808 are the only halo globulars with a {\em bimodal} HB. NGC~1261 is 
also nearby to the ``group'', it has similar metallicity and blue HB morphology 
but it doesn't fit so well into the correlations shown in Fig.~2.
According to
the method introduced by Lynden-Bell \& Lynden-Bell (1995) NGC~1851, NGC~1904, 
NGC~2298, and NGC~2808 do not appear to
have the same energy and angular momentum but they are relatively close to the
Galactic Center and the
approximation of the galactocentric velocity with the heliocentric velocity
(that is quite relevant in this kind of analysis) may not be appropriate.


\end{document}